# Demographic Transition Theory Contradicted Repeatedly by Data

Ron W Nielsen[1]

Environmental Futures Research Institute, Gold Coast Campus, Griffith University, Qld, 4222, Australia

In the absence of convincing evidence, data for Sweden and Mauritius are used in academic publications to illustrate the Demographic Transition Theory. These data are closely examined and found to be in clear contradiction of this theory. Demographic Transition Theory is also contradicted by the best available data for England. Other examples of contradicting evidence are also discussed.

**Introduction**

Historical economic growth can be studied using the Gross Domestic Product (GDP). However, to understand the time dependence of income per capita, expressed as GDP/cap, it is necessary to understand not only the economic growth, expressed in terms of the GDP, but also the growth of human population.

The latest and the most elaborate theory describing economic growth is the Unified Growth Theory (Galor, 2005, 2011). The theory, or model, describing the growth of human population is the Demographic Transition Theory (see for instance Caldwell, 1976, 2006; Casterline, 2003; Coale, 1973; Haupt & Kane, 2005; Kirk, 1996; Landry, 1934; Lee, 2003; Lehr, 2009; McFalls, 2007; Notestein, 1945; Olshansky & Ault, 1986; Olshansky, Carnes, Rogers, & Smith, 1997, 1998; Omran, 1971, 1983, 1998, 2005; Rogers & Hackenberg, 1987; Singha & Zacharia, 1984; Thompson, 1929; van de Kaa, 2008; Warf, 2010.). Both of these theories use similar approach and similar language. Both of them divide the economic growth or the growth of human population into distinctly different stages governed by distinctly different mechanisms. In particular, both of them claim an ages-long epoch of Malthusian stagnation followed by a sudden transition to a distinctly different stage, the transition described as a sudden takeoff, spurt, sprint or explosion.

A study published over 50 years ago (von Foerster, Mora & Amiot, 1960) demonstrated that the growth of the world population was hyperbolic during the AD era, showing implicitly that the epoch of stagnation did not exist and that there was no sudden transition to a new type of growth. This study has shown that the growth of human population during the AD era was following a monotonically increasing trajectory. As explained elsewhere (Nielsen, 2014), such a growth cannot be divided into distinctly different sections governed by distinctly different mechanisms of growth. A single mechanism has to be applied to the whole distribution. For reasons, which are hard to explain, this crucial publication (von Foerster, Mora & Amiot, 1960) appears to have been ignored in the demographic research.

---

[1] AKA Jan Nurzynski, r.nielsen@griffith.edu.au; ronwnielsen@gmail.com;

More recently (Nielsen, 2016a), it has been demonstrated that the growth of the world population was hyperbolic not only during the AD era, as pointed out by von Foerster, Mora and Amiot (1960) but also during the BC era. Furthermore, it has been demonstrated that there was no stagnation and consequently no transition from stagnation to a distinctly different and faster growth as claimed by the Demographic Growth Theory. This study identified only two transitions in the past but they were transitions of entirely different kind than claimed by the Demographic Transition Theory. They were transitions from hyperbolic growth to hyperbolic growth. The first transition was from a fast hyperbolic growth to a significantly slower hyperbolic growth and the second transition from a slow hyperbolic growth to a slightly faster hyperbolic growth. Thus, these two studies (Nielsen, 2016; von Foerster, Mora & Amiot, 1960) demonstrate that the Demographic Transition Theory is incorrect. Now we shall discuss additional evidence and we shall show that the Demographic Transition Theory is contradicted not only by the aggregate data describing the growth of the population but also by data describing birth and death rates.

Demographic Transition Theory has been described a ghost story (Abernethy, 1995). It should have been discarded long time ago but it is still in circulation and many a demographer would passionately defend its concepts. Abernethy wonders why this dead theory is still being resurrected and her plausible explanation is that it is because of the respect to elders. However, would elders feel happy to be so protected?

Science is full of discarded theories and explanations. This is how science works. New ideas are tried and if they do not work they are replaced by better ideas or simply abandoned. To cling to incorrect ideas just because we cannot think about something better to replace them is scientifically unjustified.

Friedman, Managing Editor of the *Population and Development Review*, claims that the Demographic Transition Theory with its "formulaic presentation of the four states" "is largely a straw man" (Friedman, 2015). This classical version of the Demographic Transition Theory is now known as the first demographic transition to which a second demographic transition has been added (Lesthaeghe, 2010, 2014; Lesthaeghe & van de Kaa, 1986; van de Kaa, 2001, 2002). The classical four stages of growth are still there even though they have no convincing support in data. "It is fair to say, that nearly all statements of a general kind about the classical - for me now the first - demographic transition, can be easily contradicted" (van de Kaa, 2002, p. 9). The classical Demographic Transition Theory appears to have been not only acknowledged but also reinforced by adding the international migration component. Kirk observed that "Demography is a science short on theory, but rich in quantification" (Kirk, 1996, p. 361) but it would be perhaps better to have science without a theory than "science" with a theory contradicted by data.

There is no science without data. In science, even the best constructed theory can be undermined and even abolished by just one contradicting evidence. It would be better to accept that it is perhaps impossible to have a general theory in the demographic research and that each case should be explained individually.

The curious feature of the Demographic Transition Theory is that *there is not a single convincing confirmation of this theory in data.* Try as we may, we shall never find data showing convincingly the four stages of growth. It is for this reason that Montgomery had to stitch the data for Sweden and Mauritius to *illustrate* this theory (Montgomery, n.d.). "I used Mauritius and added Sweden to the end of it. I smoothed the stage 1 of Mauritius a bit. It is composite more than purely conceptual" (Montgomery, 2012). It should be emphasised that his aim was not to prove this theory but only to illustrate it.

Data for Sweden are repeatedly used in support of the Demographic Transition Theory but we shall show that these data serve as an excellent illustration that the Demographic Transition Theory is contradicted by empirical evidence. Data for Mauritius are sometimes used but we shall show that they also do not support this theory. The best and the most extensive data are for England (Wrigley & Schofield, 1981). We shall demonstrate that the Demographic Transition Theory is also contradicted by these data.

It is taken for granted that the first stage, which is believed to have lasted for thousands of years, was characterised by strong fluctuations in birth and death rates but we have absolutely no data to prove it. We do not have data for death and birth rates extending over thousands of years, so in this sense at least this part of the theory is unscientific. We have no choice but to accept it by faith.

However, much more has to be accepted by faith. No-one has ever proven the existence of the first stage of growth (the epoch of stagnation) proposed by the Demographic Transition Theory. In fact, this concept is contradicted by data (Nielsen, 2013, 2016a; von Foerster, Mora & Amiot, 1960). There was no stagnation in the growth of human population but for doctrines accepted by faith, contradictions in data are routinely and promptly ignored. The only way to accept this stage of growth is by faith and by ignoring population data (Maddison, 2010; Manning, 2008; US Bureau of Census, 2016a) and their contradicting evidence, but then it is no longer science. Countless descriptions of this mythical epoch and of the mechanism of growth during that long time have to be accepted by faith.

No-one has ever proven that there was a transition from the first to the second stage. No-one has ever proven that there was population explosion at a certain time. Rapid growth of the population, interpreted as population explosion, is real but it is just the natural continuation of hyperbolic growth (Nielsen, 2014, 2016a), the type of growth, which was identified over 50 years ago (von Foerster, Mora & Amiot, 1960) but which was also conveniently ignored. The transition from the alleged first to the second stage has to be accepted by faith and by ignoring not only the evidence published over 50 years ago but also the extensive population data (Maddison, 2010; Manning, 2008; US Bureau of Census, 2015).

No-one has ever proven that the mechanisms of growth during the alleged first and second stages were different. No-one has ever proven that the Industrial Revolution boosted the growth of human population. All these concepts and more have to be accepted by faith supported perhaps occasionally by the misinterpretation of selected data.

It is believed that strong fluctuations in birth and death rates are reflected in fluctuations in the size of the population. These assumed fluctuations, described often as Malthusian oscillations, have been extensively discussed in peer-reviewed literature but no-one cared to check whether fluctuations in birth and death rates have any influence on the growth of human population. We shall demonstrate that these *fluctuations have absolutely no impact on the growth of human population*.

It is believed that the growth of the population was stagnant for thousands of years and that it was characterised by random variations. According to this belief, there were periods of time when the population *did not grow at all* and that any gains in the growth of human population made over decades were *wiped out in one or two years* (van de Kaa, 2008). Such confident declarations are inaccurate and misleading. They might apply to some local populations, sometimes, but they certainly do not apply to the growth of the world population (Nielsen, 2013, 2016a). This claim is also not supported by the regional population data (Maddison, 2010).

Normally, in any scientific investigation, empirical evidence such as published over 50 years ago (von Foerster, Mora & Amiot, 1960) would have been further investigated. Why was it ignored in the demographic research? This early observation is now convincingly confirmed (Nielsen, 2016a) by new data (Maddison, 2010; Manning, 2008; US Bureau of Census, 2015). The growth of the population in the past was hyperbolic. It was slow but it was not stagnant or random. *The first stage proposed by the Demographic Transition Theory did not exist and there was no transition from stagnation to growth.*

**Demographic Transition Theory**

We have already mentioned certain features of the Demographic Transition Theory but in order to understand the discussed examples for Sweden, Mauritius and England we shall now present its brief outline. Its general concepts are illustrated in Figure 1.

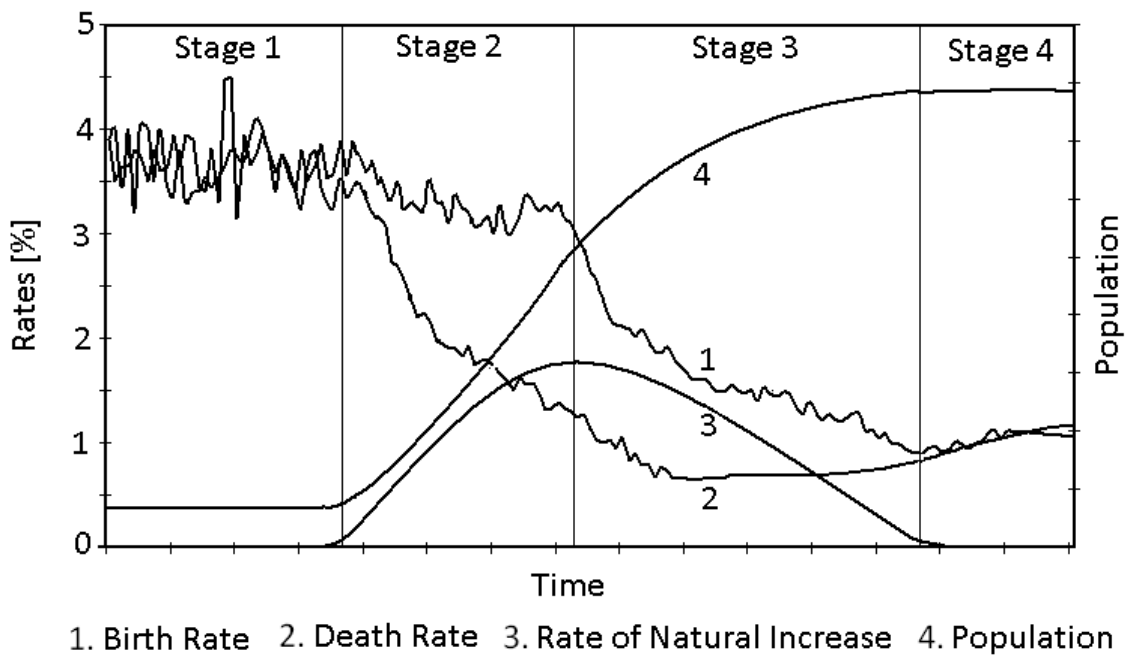

**Figure 1.** Fundamental concepts of the Demographic Transition Theory based on the illustrations presented by Montgomery (n.d.) and by van de Kaa (2001, 2002).

Demographic Transition Theory describes changes in birth and death rates, in the size of the population and in the rate of natural increase. According to this theory, changes in socio-economic conditions lead to transitions in death and birth rates, which in turn are reflected in the growth of human population (see for instance Caldwell, 1976, 2006; Casterline, 2003; Coale, 1973; Haupt & Kane, 2005; Kirk, 1996; Landry, 1934; Lee, 2003; Lehr, 2009; McFalls, 2007; Notestein, 1945; Olshansky & Ault, 1986; Olshansky, Carnes, Rogers, & Smith, 1997, 1998; Omran, 1971, 1983, 1998, 2005; Rogers & Hackenberg, 1987; Singha & Zacharia, 1984; Thompson, 1929; van de Kaa, 2008; Warf, 2010). These transitions are supposed to have been taking place in four fundamental stages, to which other stages could be added.

Stage 1 is supposed to have been the pre-industrial stage of stagnation; Stage 2 is supposed to represent the post-industrial stage of explosion; Stage 3 is the stage of the slowing-down growth; and Stage 4 is the stage of a stable size of the population. The number of stages can be extended to five (Haupt & Kane, 2005; Olshansky, Carnes, Rogers, & Smith, 1998; van de Kaa, 2008) or maybe even to six (Myrskyla, Kohler & Billari, 2009).

The theory was proposed in its inchoate form in 1929 (Thompson, 1929) but the word "transition" was not used until 1934 (Landry 1934). The first clear outline of this theory is attributed to Notestein (1945). Its fundamental concepts illustrated in Figure 1 are based on the illustration prepared by Montgomery (n.d) and by van de Kaa (2001, 2002).

**Stage 1** is claimed to have **"prevailed since time immemorial"** (Komlos, 2000, p. 320), i.e. for many thousands of years. The characteristic feature of this stage is the high birth and death rates fluctuating around the same constant value and producing a stagnant state of growth. The size of the population remained approximately constant and the rate of natural increase approximately zero. This stage is described as the Preindustrial Age, the Preindustrial Society, the Malthusian Regime, the Epoch of Malthusian Stagnation, the Pre-Demographic Transition Stage and the Age of Pestilence and Famine. Living conditions during that long time are claimed to have been characterised by poor health care, poor hygiene, "inadequate diets, as well as unsanitary drinking water and bacterial diseases" (Warf, 2010, p. 708). During this stage, there was a continuing struggle for survival and the growth of the population was "fluctuating around zero" (Warf, 2010, p. 708).

**Stage 2** is supposed to have been dramatically different. It was the stage of population explosion, usually linked with the Industrial Revolution, the stage of transition from ages-long stagnation to a rapid growth of the population. The rate of natural increase is supposed to have started to increase rapidly and the size of the population exploded. This stage is described as the Early Industrial Society, the Early Industrial Age, the Post-Malthusian Regime, the Early-Demographic Transition and the Age of Receding Pandemics. The transition from Stage 1 to Stage 2 is described as the Escape from the Malthusian Trap, the Great Escape and as the population explosion.

The characteristic feature of this stage is supposed to have been the rapidly declining death rate described as the *mortality transition*, allegedly caused by the generally improving living conditions reflected in a substantially better health care, better hygiene, better access to clean water, improved sanitation and increased food production (Chrispeels and Sadava 1994; Galor and Weil, 2000; Thomlinson, 1965). These postulated new growth-promoting forces "ignited a population explosion" (McFalls 2007, p. 26). Another characteristic feature of this stage is the continuing high birth rate over a certain time followed by its gradual decline.

**Stage 3** is the stage of the slowing down growth and is described as the Mature Industrial Age, the Late Industrial Society, the Modern Growth Regime, the Stage of the Late Demographic Transition, or the Stage of Degenerative and Man-made Diseases. The difference between the mechanism of growth in Stages 2 and 3, is explained by a change in personal preferences prompted by such factors as women joining work force, better education, the availability of contraceptives and by the general tendency to have smaller number of children in order to improve the standard of living.

**Stage 4** is the stage of a stable size of the population and is described as the Post-industrial Society, the Post-industrial Age, the Age of Delayed Degenerative Diseases, the Post-Demographic Transition Stage or the Stage of Invincibility. This stage is characterised by a close balance between birth and death rates, similar to the balance claimed for the Stage 1, but now both rates are low. Low birth rate is explained by personal preferences of replacing quantity by quality. The impact of infectious diseases during this stage is claimed to be low and to be replaced by harmful changes in the lifestyle. Mortality is now "associated with smoking and obesity, as well as, to a lesser extent, car accidents, suicides, and homicides" (Warf, 2010, p. 710).

We shall now examine the data for Sweden, Mauritius and England and we shall show that they are in contradiction with the Demographic Transition Theory, but in perfect harmony with other contradicting evidence (Nielsen, 2013, 2016a; von Foerster, Mora & Amiot, 1960). The evidence is already strong. Demographic Transition Theory has no place in science.

In comparing this theory with empirical evidence it is essential to understand the characteristic features of the alleged Stage 1 and of the transition to Stage 2.

1. The alleged Stage 1 should be characterised by strong fluctuations in birth and death rates.
2. On average, birth and death rates should be high and nearly constant.
3. The gap between the fluctuating birth and death rates during this first stage should be on average zero.
4. There should be convincing evidence of stagnation in the growth of the population during the alleged Stage 1.
5. There should be a clear and convincing transition from Stage 1 to Stage 2, marked by a clear change in the pattern of growth of human population, *from stagnation to growth*, so clear that it could be described as a takeoff, spurt, or explosion.
6. The transition should be marked by a clear change in the pattern of birth and death rates. On average, death rates should start to decrease, while the birth rates should, for a certain limited time, remain constant and then they should also start to decrease.
7. The gap between birth and death rates should be progressively getting wider from approximately zero to a certain maximum value, which would mark the beginning of Stage 3.

**Examination of data for Sweden**

The data for Sweden (Statistics Sweden, 1999), used repeatedly in support of the Demographic Transition Theory, are displayed in Figure 2.

These data appear to be in support of the four stages of growth (*cf* Figure 1): Stage 1 characterised by large, nearly constant and strongly fluctuating birth and death rates; Stage 2 characterised by a widening gap between the average values of birth and death rates; Stage 3 characterised by a decreasing difference between the birth and death rates; and Stage 4 characterised by low and nearly equal birth and death rates. These data show also the gradually decreasing fluctuations in birth and death rates.

However, what should notice immediately is that birth and death rates in the alleged Stage 1 do not fluctuate around the same constant value.

In order to produce a stagnant state of growth, birth and death rates have to vary around the same constant value. It is essential for the difference between them to be on average zero. It could vary between negative or positive values but it should not be on average larger than zero.

Data for Sweden should have never been used to illustrate the Demographic Transition Theory. Likewise, data for death rates *or* birth rates should never be used to test the Demographic Transition Theory. They should be used *together* because the Demographic Transition Theory describes how *both of them* should behave.

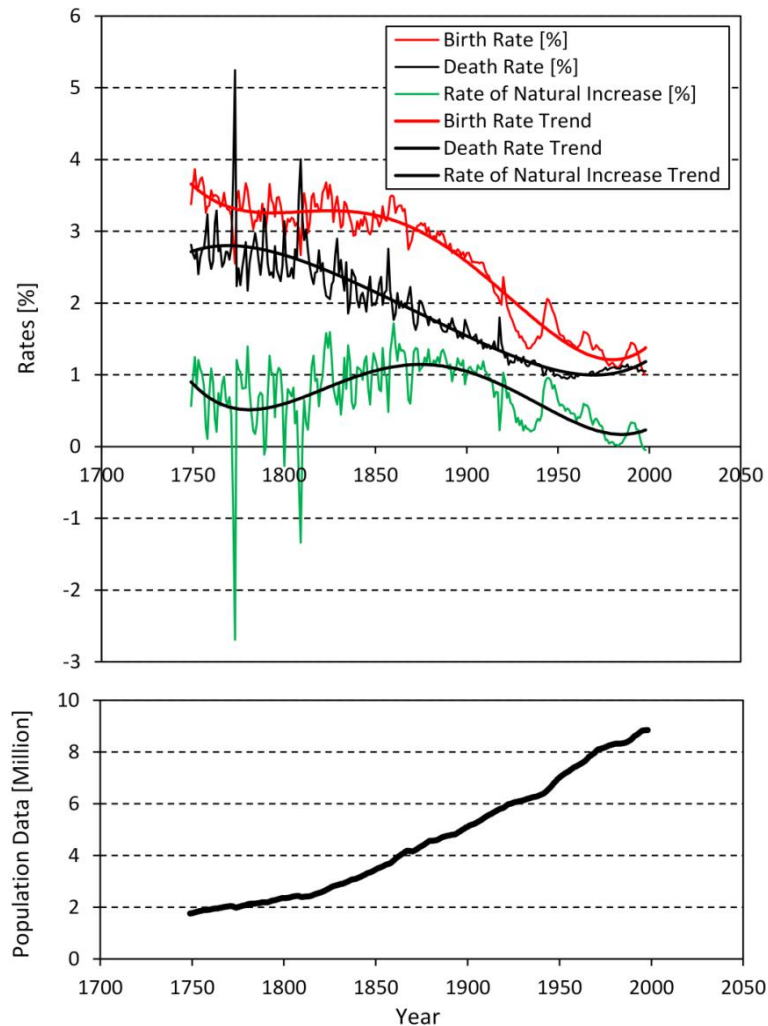

**Figure 2.** Demographic Transition Theory is contradicted by the data for Sweden (Statistics Sweden, 1999). *The four stages of growth did not exist*. There was no stagnation and no explosion, takeoff or spurt in the growth of the population. The data also show that even large fluctuations in birth and death rates have no impact on the growth of human population.

In the same source (Statistics Sweden, 1999) there are also aggregate data describing the growth of the population in Sweden, shown in the lower section of Figure 2. These data clearly demonstrate that the four stages of growth did not exist. They should have never been ignored in testing the Demographic Transition Theory.

Data for Sweden should have never been used to illustrate the validity of the Demographic Transition Theory because such illustrations are incorrect and misleading. When used in classrooms or lecture rooms, they do not teach science. When used in academic publications in support of the Demographic Transition Theory they propagate unscientific and incorrect information.

Death rate shown in the upper section of Figure 2 is decreasing in the apparent agreement with Stage 2 but it was also decreasing in the apparent Stage 1. There was no clear mortality transition, which could be claimed as marking the change from Stage 1 to Stage 2.

Consequently, the apparent Stage 2 cannot be identified as Stage 2, which puts in questions other apparent stages.

The widening gap during the apparent Stage 2 is only slightly larger than the gap during the apparent Stage 1. Such a small change could not have produced a desired transition from a stagnant growth during the alleged Stage 1 to an explosive growth during the apparent Stage 2. In fact, the wide gap between birth and death rates during the alleged Stage 1 is obviously so large that there must have been no stagnation during this stage but a steadily-increasing growth of the population, and indeed this expectation is confirmed by the aggregate data describing the growth of the population in Sweden and shown in the lower part of Figure 2.

The disagreement between data and the Demographic Transition Theory is made even clearer if we look at the growth of human population in Sweden. They show clearly that *the four stages of growth did not exist*. Demographic Transition Theory neither describes nor explains the growth of human population in Sweden and is in gross disagreement with data

The data also show that even violent fluctuations in birth and death rates and the resulting fluctuations in the rate of natural increase have no impact on the growth of human population. The fluctuations in birth and death rates did not produce the normally expected Malthusian oscillations in the growth of human population.

A study of such fluctuations might be interesting for another reason but it has no bearing on explaining the mechanism of growth of human population. If we look at Figure 2, we can see that some points for the rate of natural increase are located far from the prevailing trend and yet even such large fluctuation had no noticeable effect on the recorded size of the population.

Summary of the contradicting evidence:

1. Contrary to the Demographic Transition Theory, the gap between birth and death rates during the alleged Stage 1 is not close to zero.
2. Such a wide gap cannot produce a stagnant state of growth characterised by a zero rate of natural increase, and indeed the data show that the rate of natural increase during this alleged Stage 1 was not zero.
3. The gap between birth and death rates during the alleged Stage 2 is only slightly larger than during the alleged Stage 1.
4. Such a difference is the size of the gap cannot produce the population explosion, and indeed there was no population explosion in Sweden during the displayed time.
5. "Mortality transition" (the decreasing death rate) commenced during the alleged Stage 1 and consequently, the alleged Stage 1 is not Stage 1.
6. Population data demonstrate that the four stages of growth did not exist. They show that there was a steadily increasing growth of the population.
7. Demographic Transition Theory is contradicted by the data for Sweden.

**Examination of data for Mauritius**

Data for Mauritius (Lehmeyer, 2004; Mauritius, 2015; Statistics Mauritius, 2014; UN, 2013) are shown in Figure 3.

Using the data for Mauritius in support of the Demographic Transition Theory (e.g. Lutz & Qiang, 2002) is surprising, because the population in Mauritius represents a minute fraction of the world population, and thus these data can be hardly considered as representing a typical patterns of birth and death rates. Furthermore, these data are poorly documented and it is uncertain, which areas were included in the population surveys.

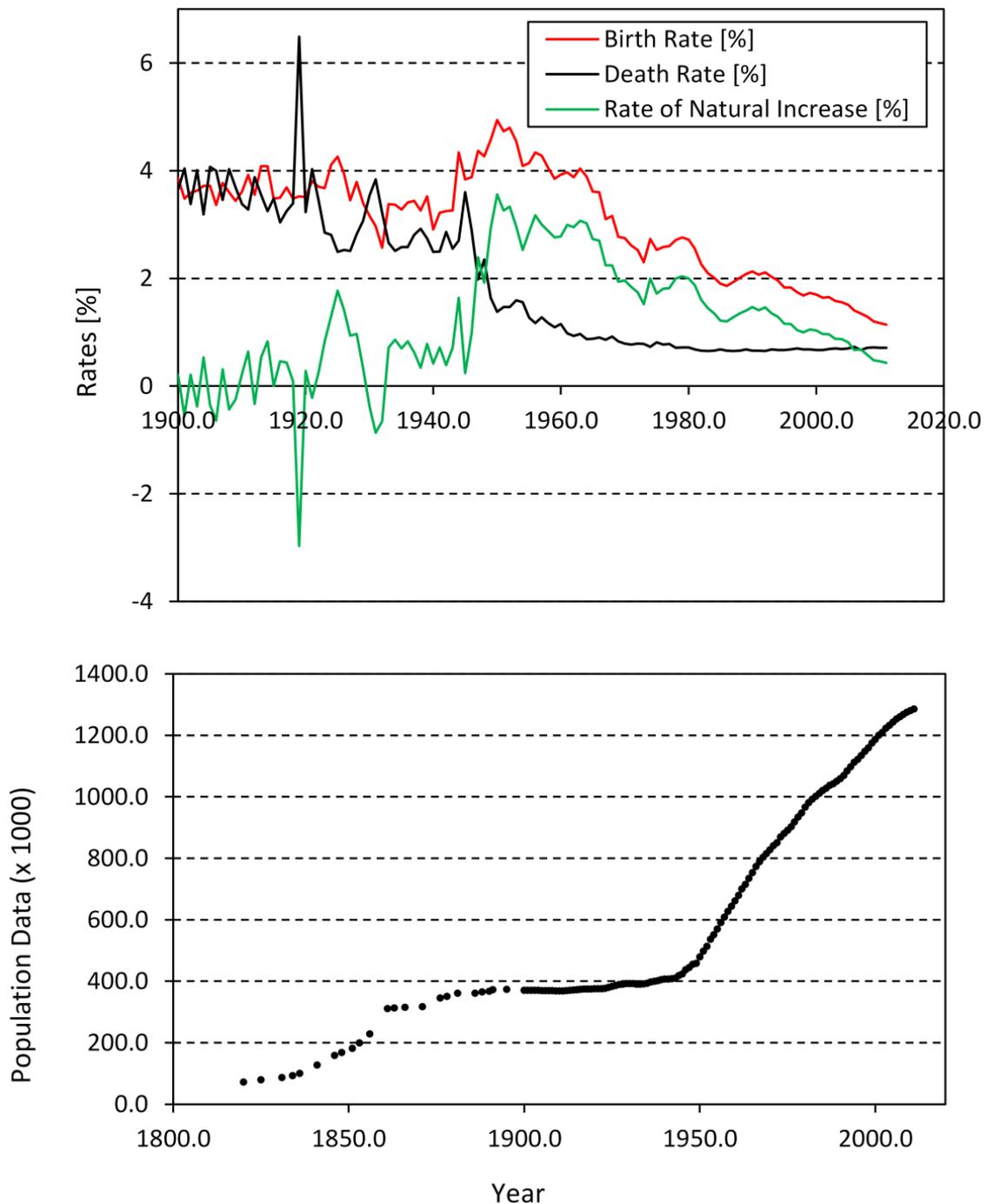

**Figure 3.** Demographic Transition Theory is contradicted by the data for Mauritius (Lehmeyer, 2004; Mauritius, 2015; Statistics Mauritius, 2014; UN, 2013). The apparent Stage 1 is not Stage 1 because it was preceded by a fast growth of the population. Please notice that the time scales for the two diagrams are not the same.

The data describing birth and death rates, shown in the upper section of Figure 3, appear to be supporting the Demographic Transition Theory. Birth and death rates are at first high and they appear to be fluctuating around the same constant value, suggesting Stage 1 of growth.

There is also a clear mortality transition at a certain time marked by the rapidly decreasing death rate, accompanied by an increasing gap between birth and death rates, in good agreement with the pattern expected for Stage 2, characterised by a transition from stagnation to an explosive growth of the population. Gradually, the gap between birth and death rates narrows suggesting Stage 3 with a possibility of developing into Stage 4.

However, this apparent agreement with the theory becomes questionable when we look at the time scale. The "epoch of stagnation" as indicated by the merging birth and death rates lasted for only around 20 years. We could, perhaps, extend it to 40 years but we can see that the gap between birth rates started to increase from around 1920. The "epoch" is probably nothing more than a temporary delay in the growth of the population.

One of the fundamental principles of scientific investigation is that no relevant data should be ignored. Consequently, in order to understand the patterns displayed by birth and death rates we have to include also the data describing the growth of the population. These data are shown in the lower part of Figure 3 and they now make it perfectly clear that they do not support the Demographic Transition Theory, because the population was increasing before the apparent Stage 1. Consequently, the apparent Stage 1 is not Stage 1, which means that the apparent Stage 2 is not Stage 2. The whole pattern of growth is incompatible with the Demographic Transition Theory. The growth of the population in Mauritius was increasing, sometimes faster and sometimes slower, in complete disagreement with the Demographic Transition Theory.

The data show that over the displayed time the growth of the population was at first slow, then fast, slowing down, slow, fast, and slowing down again. Data for Mauritius demonstrate that there were more demographic transitions than claimed by the Demographic Transition Theory.

Summary of the contradicting evidence:

1. While the gap between birth and death rates is close to zero as required by the Demographic Transition Theory for the Stage 1, the empirical evidence indicates that the stagnant state of growth lasted for only about 20 years or at best for only 40 years. The required evidence should be for at least a few hundred years, but in principle it should be for thousands of years.

2. The apparent Stage 2 looks like Stage 2 but this interpretation is contradicted by the population data showing that a similar stage of a fast growth was before the apparent Stage 1

3. Population data show that there were three, maybe even four, stages of growth during the displayed short time but these stages have nothing to do with the Demographic Transition Theory.

4. Demographic Transition Theory is contradicted by the data for Mauritius.

**Examination of data for England**

Probably the best, the most reliable and the most extensive demographic data we might ever expect to have are for England (Wrigley & Schofield, 1981) between 1541 and 1871. These data are important not only because of their high accuracy but also because they extend into the time well before of the Industrial Revolution, dated between 1760 and 1840 (Floud & McCloskey, 1994). Furthermore, it is important that these data are for England, where the impacts of the Industrial Revolution on the growth of human population should be strong and clear.

It is here, in England, that we should expect a clear confirmation of a change from high birth and death rates fluctuating around the same constant value to a new pattern characterised by a rapidly widening gap between these two quantities, indicating a clear transition from stagnation to population explosion. It is here, in England, that we should be able to see a clear *correlation* between the Industrial Revolution and the morality transition (the decreasing death rate); the clear confirmation of the beneficial effects of modern progress; the clear evidence of a dramatic escape from the Malthusian Trap; the dramatic transition from Malthusian stagnation (marked by a stagnant stage of growth characterised by Malthusian oscillations) to a rapid and sustained growth of human population.

Birth and death rates, together with the corresponding rate of natural increase in England are shown in Figure 4. The time-dependent patterns are entirely different than claimed by the Demographic Transition Theory (*cf* Figure 1).

Birth and death rates were always high – before, during and after the Industrial Revolution. They were also not fluctuating around a common constant value before the Industrial Revolution and there was no morality transition coinciding with this event. In fact, Industrial Revolution had no impact on the time-dependent distributions of birth and death rates. It is as if this crucial development, which was supposed to have had such a dramatic impact on the growth of human population had never happened.

The data for birth and death rates and for the population, shown in Figures 4 and in the upper panel of Figure 5, respectively, are at yearly intervals. However, while the data for birth and death rates and for the corresponding rate of natural increase show strong fluctuations, the data for the growth of the population does not show even a slightest effect of these fluctuations. *The growth of the population is immune to the fluctuations in birth and death rates.*

"These models of Malthusian oscillations, although elegant and intriguing, must be viewed as quite speculative in their application to any actual populations" (Lee, 1997, p. 1097). Indeed, their presence is contradicted by the data for England, Sweden and Mauritius as well as by the analyses of the world population data (Nielsen, 2013, 2016a).

We have demonstrated that the fluctuations in birth and death rates are not reflected in the growth of human population. However, we can reverse our investigation and ask whether the fluctuations in birth and death rates can generate fluctuations in the *calculated* growth of human population. Suppose we use the empirically-determined birth and death rates or the rate of the natural increase representing the difference between the birth and death rates, and suppose that we use these rates to *calculate* the size of the population, will they produce the fluctuations *in the calculated distribution*?

In order to answer this question we have carried out numerical integration of the following differential equation:

$$\frac{1}{S(t)}\frac{dS(t)}{dt} = R_e(t) \qquad (1)$$

where $S(t)$ is the *calculated* size of human population and $R_e(t)$ is the empirically-determined, and *fluctuating*, rate of natural increase shown in Figure 4 and calculated using the empirically-determined, and *fluctuating*, birth and death rates. Migration rates are relatively small (Wrigley & Schofield, 1981) and can be neglected. However, if the calculated distribution of the size of the population is not going to agree with population data, they will have to be included.

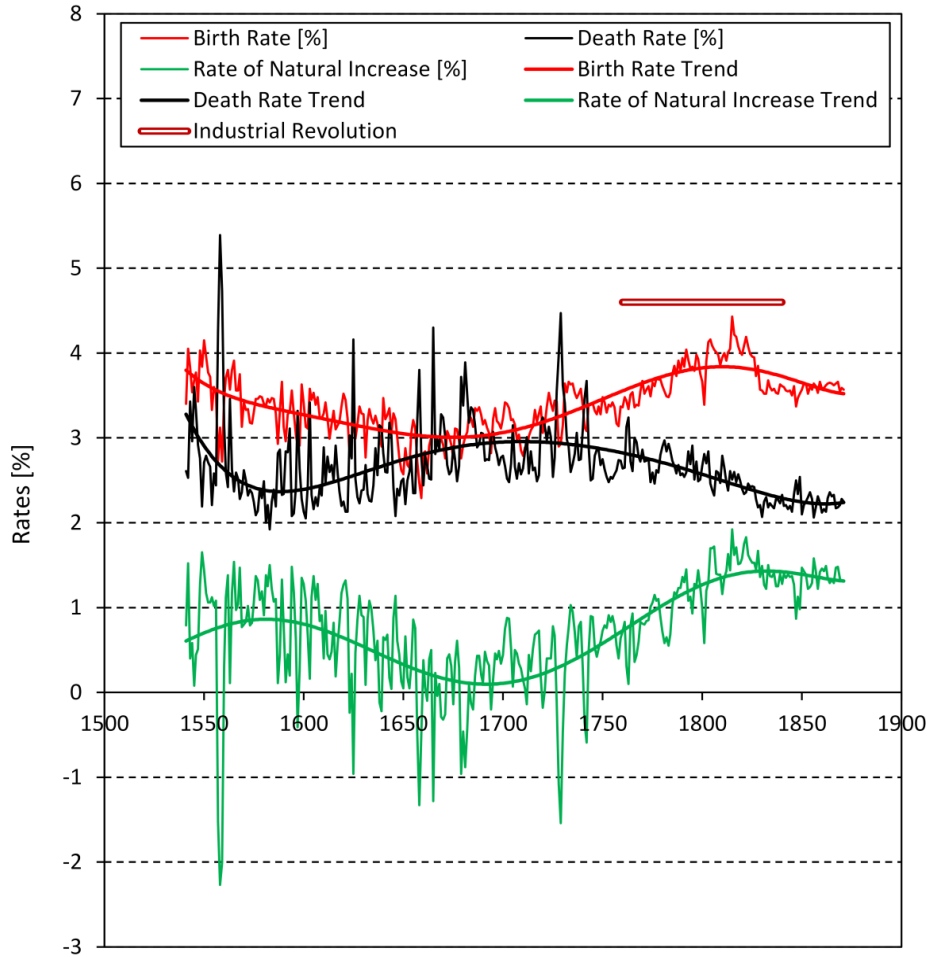

**Figure 4.** Demographic Transition Theory is contradicted by the data for England (Wrigley & Schofield, 1981). The patterns claimed for the first two stages of growth are absent. Demographic Transition Theory presents a completely different story than the data.

Results of these numerical calculations are shown in the lower part of Figure 5. The calculated curve is displayed in steps of one year but it follows that data so closely that in order to see any possible fluctuations we had to show data at 10-year intervals. The fluctuating birth and death rates or the corresponding fluctuating rate of natural increase do not produce even the slightest fluctuations in the calculated distribution describing the growth of the population.

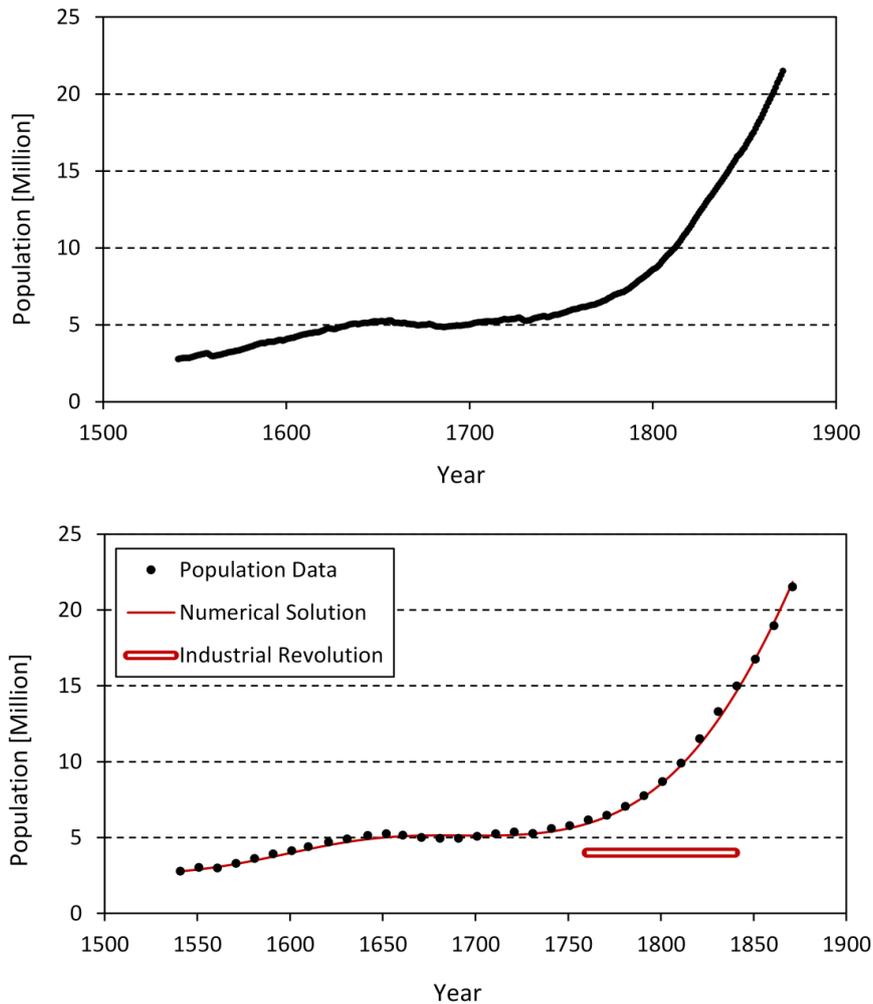

**Figure 5**. Growth of human population in England (Wrigley & Schofield, 1981). Fluctuations in birth and death rates shown in Figure 4 had no impact on the growth of the population. The typical pattern of stagnation followed by explosion is not confirmed by data. The population was increasing well before the onset of the Industrial Revolution. After a short delay, the population started to increase again but the onset of this new growth was also before the Industrial Revolution.

The growth of the population shown in Figure 5 does not display the expected pattern of stagnation followed by explosion claimed by the Demographic Transition Theory. There was a steady growth of the population well before the Industrial Revolution. This growth was briefly interrupted but it was resumed again around 1700. The growth of the population in England is not correlated with the Industrial Revolution. There is no indication of prolonged Malthusian stagnation, no evidence of Malthusian oscillations, no clear evidence of the existence of stage one and no transition to a new stage. It is just a growth, which was increasing, halted for a while and started to increase again. Demographic Transition Theory is contradicted by data.

The growth of human population can be also studied using the reciprocal values of data, $1/S(t)$. Such a study gives a new insight into the interpretation of data. This method has been discussed elsewhere (Nielsen, 2014).

Reciprocal values of the size of human population in England and their absolute gradient, calculated directly from data and interpolated, are shown in Figure 6.

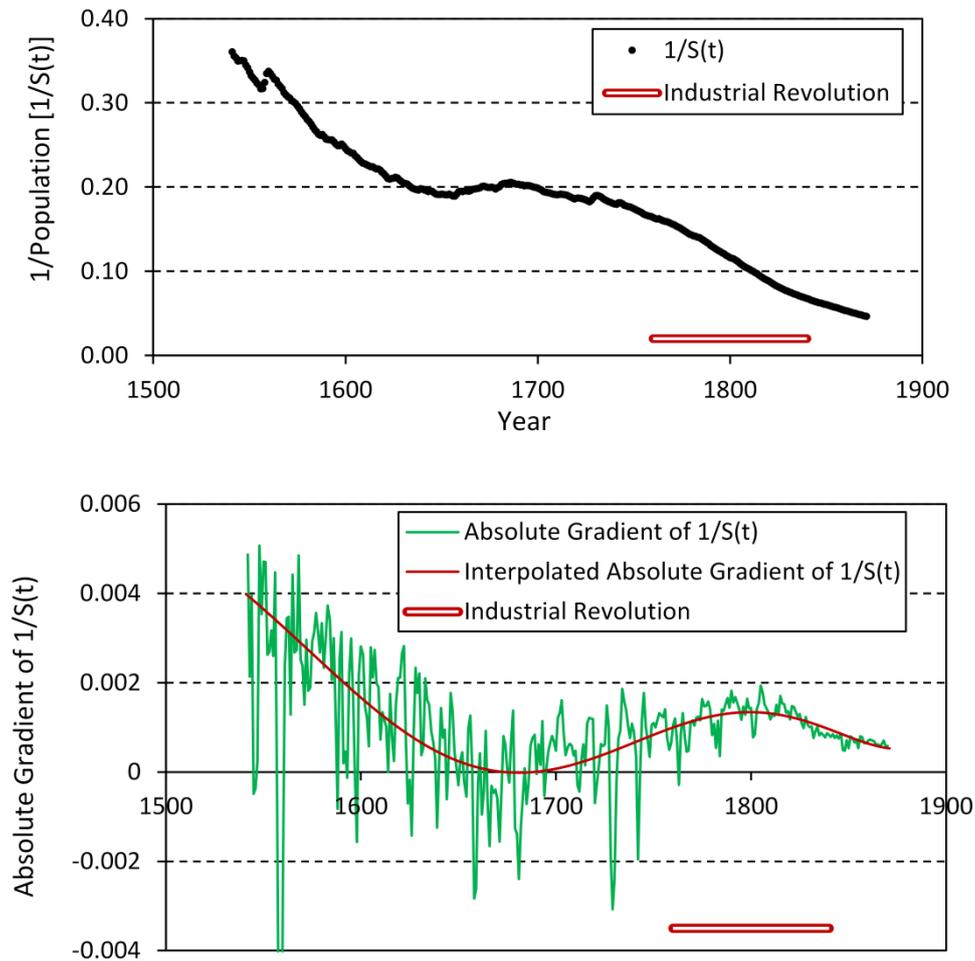

**Figure 6.** Reciprocal values, $1/S(t)$, of the size of the population and their absolute gradient calculated directly from data and interpolated. There was no stagnation before the Industrial Revolution and no boosting of growth by the Industrial Revolution. On the contrary, the Industrial Revolution coincides with the slowing-down growth as indicated by the maximum in the interpolated gradient.

The deceasing reciprocal values, $1/S(t)$, of the size of the population indicate an increasing growth, and *vice versa*. The top section of Figure 6 shows that the population in England was steadily increasing well before the Industrial Revolution, as indicated by the steadily-decreasing reciprocal values. After only a brief interruption, the size of the population in England continued to increase, confirming the pattern of growth shown in Figure 5.

There was no stagnation that could be identified as Stage 1 proposed by the Demographic Transition Theory. This stage did not exist. In its place there was, in general, a steadily-increasing growth. The data show a temporary distortion of this trajectory but the general pattern was a continuing increase of the population. Furthermore, the data show no correlation of the growth of the population with the Industrial Revolution. There was no transition to a distinctly new stage. This pattern of growth is in contradiction with the pattern proposed by the Demographic Transition Theory.

The absolute gradient of the reciprocal values, $1/S(t)$, of the size of the population is also a convenient indicator allowing for detecting whether the growth was accelerating or decelerating. The decreasing absolute gradient indicates a slowing-down growth while the increasing gradient indicates an acceleration.

The absolute gradient of the $1/S(t)$ data is shown in the lower part of Figure 6. If we compare the upper and the lower sections of this figure we can see that the growth of the population in England was steadily increasing, as indicated by the decreasing reciprocal values $1/S(t)$, but it was gradually getting slower, as indicated by the decreasing absolute gradient of the reciprocal values. After a short period of instability, the growth of human population in England started to increase again from around 1690 and was accelerating, as indicated by the downward bending of the reciprocal trajectory and by its increasing absolute gradient. The onset of this new growth occurred about 70 years before the onset of the Industrial Revolution. Contrary to the general beliefs, *the Industrial Revolution did not boost the growth of human population in England* where its impacts should be stronger than anywhere else.

After a certain time, the acceleration of the growth of human population started to grow weaker, as indicated by the gradient approaching its maximum value. The absolute gradient of the reciprocal values reached its maximum around 1800 or right in the middle of the Industrial Revolution and then started to decrease. The growth of human population started to decelerate.

If we wanted to claim a cause-effect link between the Industrial Revolution and the growth of human population in England we could conclude that the Industrial Revolutions slowed down the growth of the population and diverted it to a slower trend as indicated by the decreasing absolute gradient of the reciprocal values. However, more plausible conclusion is that the *Industrial Revolution had no impact on the growth of human population*. The two processes were totally independent and it is incorrect to link them by any cause-effect properties. The growth of human population in England must have been prompted by different forces than the forces associated with the Industrial Revolution and with the numerous random forces repeatedly proposed to explain the epoch of stagnation, which did not exist.

Summary of the contradicting evidence:

1. The time dependence of birth and death rates in England between 1541 and 1871 (Wrigley & Schofield, 1981) is in contradiction with the first two stages of growth claimed by the Demographic Transition Theory. The pattern of the fluctuating birth and death rates around a common high constant value followed by a clear transition to a new stage around the time of the Industrial Revolution is contradicted by data.

2. The first stage of growth proposed by the Demographic Transition Theory did not exist.

3. The data show not just one mortality transition (decreasing death rate) as claimed by the Demographic Transition Theory, but two, both of them beginning well before the onset of the Industrial Revolution.

4. There is no positive correlation between the Industrial Revolution and the time dependence of birth and death rates.

5. Data for England show that there were more demographic transitions than can be accounted for by the Demographic Transition Theory.

6. With the exception of a minor delay between around 1656 and 1682, the growth of human population in England was steadily increasing.

7. Reciprocal values of data for the size of human population also confirm that Industrial Revolution had no impact on the growth of the population in England, where it should have been stronger than anywhere else

8. Rather than being boosted by the Industrial Revolution, the growth of the population in England started to be diverted to a slower trajectory from around 1800.

9. Demographic Transition Theory is contradicted by the data for England between 1541 and 1871.

**Summary and conclusions**

Data for Sweden, used repeatedly in support of the Demographic Transition Theory, are shown to be in its direct contradiction. They show that the four stages of growth claimed by the Demographic Transition Theory did not exist. There was no stagnation (no Stage 1) and no transition to a new stage (Stage 2) claimed by the Demographic Transition Theory. There was no population explosion and no transitions to stages three and four. There was just a steadily-increasing, single-stage, growth of the population. The gap between death and birth rates in the apparent Stage 1 was large and there was no dramatic change in its size during the usually claimed but non-existent transition from Stage 1 to Stage 2.

The data for Mauritius, used sometimes in support of the Demographic Transition Theory (e.g. Lutz & Qiang, 2002) also show a clear disagreement with this theory. The apparent Stage 1 suggested by the birth and death rates, even if accepted, lasted for only a few decades. However, when aggregate data are included, they show that the apparent Stage 1 was not Stage 1 because it was not preceded by stagnation but by a steadily increasing growth of the population. The pattern of growth of human population does not fit into the pattern claimed by the Demographic Transition Theory

The exceptionally good data for England, 1541-1871, are also in contradiction with the Demographic Transition Theory. The expected stages in birth and death rates are not confirmed by the data. There were two mortality transitions during that time, both commencing well before the onset of the Industrial Revolution. There is no correlation between the Industrial Revolution and the time-dependence of the birth and death rates in England. There was no stagnation followed by population explosion.

Industrial Revolution did not boost the growth of human population in England. On the contrary, the data show that from around 1800 the growth of the population in England started to be slowing down. Consequently, if we want to link the Industrial Revolution with the growth of the population we would have to conclude that the Industrial Revolution slowed down the growth of the population. However, more plausible conclusion is that the two processes were totally independent. The data indicate that it is incorrect to use the Industrial Revolution to explain the mechanism of growth of human population, even in England, the centre of this revolution.

While the data for Sweden show a steady growth of the population without any signs of four stages claimed by the Demographic Transition Theory, the data for Mauritius and England demonstrate that there were more stages than one can account for by using the Demographic Transition Theory.

The study presented here also shows that even large fluctuations in birth and death rates have no impact on the growth of human population. It is, therefore, incorrect to imagine that fluctuations in birth and death rates can produce Malthusian oscillations in the size of the population. Furthermore, moderate variations in the growth rate or the rate of natural increase can, at best, create only small and negligible variations in the growth of population.

A study published over 50 years ago (von Foerster, Mora & Amiot, 1960) demonstrated that the growth of human population was hyperbolic. In science, even one contradicting evidence is sufficient to show that a contradicted theory is incorrect. Now we have more extensive sources of data (Maddison, 2010; Manning, 2008; US Bureau of Census, 2015). They all show that there was no stagnation in the growth of human population (Nielsen, 2013, 2016a). They show clearly that the Stage 1 claimed by the Demographic Transition Theory did not exist and that there was no transition to the alleged Stage 2. They show that the Demographic Transition Theory is contradicted by data.

Demographic Transition Theory has a strong link with the Unified Growth Theory (Galor, 2005, 2010), which also claims, incorrectly, the existence of the epoch of stagnation and a dramatic transition to a new stage of economic growth described repeatedly as takeoff. A study of the income per capita (GDP/cap) combines the study of the economic growth, as expressed by the GDP, and the study of the growth of the population.

The time distribution of the historical GDP/cap values is claimed in the Unified Growth Theory to be made of a prolonged stagnation followed by a sudden takeoff in much the same way as the Demographic Transition Theory claims that the growth of human population can be represented by a prolonged stage of stagnation followed by a sudden explosion. Both interpretations are incorrect and both of them are based by illusions reinforced by the incorrect interpretations of hyperbolic growth.

The growth of the population was slow over a long time and fast over a short time but it was slow because it was hyperbolic and fast because it was hyperbolic. It was a monotonically-increasing hyperbolic distribution (Nielsen, 2016a; von Foerster, Mora & Amiot, 1960). Economic growth, whether expressed in terms of the GDP or GDP/cap, was slow over a long time and fast over a short time but it was slow because it was hyperbolic and fast because it was hyperbolic (Nielsen, 2014, 2015b). There was no stagnation and no sudden takeoff.

Demographic Transition Theory is incorrect and the only way to accept it is by ignoring the repeatedly contradicting empirical evidence and by placing full trust in stories based largely on creative imagination perfected by accretion over many years and by many people, each new imagined explanation or concept creating new ideas and all growing into the established knowledge in demography.

The Demographic Transition Theory (or Model) has been a ghost story for at least 20 years (Abernethy, 1995) and it is not clear why its concepts have not been abandoned long time ago. It would be probably better to accept openly and clearly that each case should be studied individually and that it is not necessary to reconcile them with some kind of a master theory, which at present does not exist.

Scientific principles of investigation can be used even in the absence of an all-encompassing theory, and the fundamental principle is to refrain from ignoring any relevant data particularly if they contradict the accepted interpretations. It appears that the continuing use of the Demographic Transition Theory makes the demographic research unscientific because by now and over many years this field of research evolved into a strong system of concepts many of which can be accepted only by faith.

There is also another serious problem with the continuing toleration of this theory. Demographers might be aware of the fundamental problems associated with Demographic Transition Theory. However many teachers, lecturers and university professors might be less informed. They accept it as scientific and they teach it to younger generations, who accept this theory as presented to them believing that they learn science.

For instance, quite recently, Thompson and Roberge (2015) published an article in which they present a diagram showing the four stages of growth proposed by the Demographic Transition Theory. They show how to help students to unpack "this rich display of information" (p. 254) without being aware that they are helping to unpack this rich source of *misinformation*. It would be more useful to teach students why the diagram they see is a misleading source of misinformation. It is a fiction story, a ghost story, presented as science, but teachers might not be aware of the problems permeating the corridors of science.

Correct understanding of the growth of human population is important but the misleading information presented by the Demographic Transition Theory is seriously harmful because this theory does not explain the growth of human population but presents concepts and explanations, which when closely examined are contradicted by empirical evidence. It is better to have no theory than a misleading theory.